\documentclass{article}

\PassOptionsToPackage{numbers, square}{natbib}

\usepackage[preprint]{neurips_2024}

\usepackage[utf8]{inputenc}      
\usepackage[T1]{fontenc}         
\usepackage[hidelinks]{hyperref} 
\usepackage{url}                 
\usepackage{booktabs}            
\usepackage{amsfonts}            
\usepackage{nicefrac}            
\usepackage{microtype}           
\usepackage[dvipsnames]{xcolor}  
\usepackage{amsmath} 
\usepackage{multicol,multirow}
\usepackage{graphicx}
\usepackage{algorithm}
\usepackage{algpseudocode}
\usepackage{setspace}
\newcommand{\mycomment}[1]{}

\title{Out of Many, One: Designing and Scaffolding Proteins at the Scale of the Structural Universe with Genie 2}

\author{
    Yeqing Lin\textsuperscript{1}, Minji Lee\textsuperscript{1}, Zhao Zhang\textsuperscript{2}, Mohammed AlQuraishi\textsuperscript{1}\\
    \textsuperscript{1}Department of Computer Science, Department of Systems Biology, Columbia University \\
    \textsuperscript{2}Department of Electrical and Computer Engineering, Rutgers University\\
    \texttt{\{yeqing.lin, minji.lee, m.alquraishi\}@columbia.edu}
}

\begin{document}

\maketitle

\begin{abstract}
  Protein diffusion models have emerged as a promising approach for protein design. One such pioneering model is Genie, a method that asymmetrically represents protein structures during the forward and backward processes, using simple Gaussian noising for the former and expressive SE(3)-equivariant attention for the latter. In this work we introduce Genie 2, extending Genie to capture a larger and more diverse protein structure space through architectural innovations and massive data augmentation. Genie 2 adds motif scaffolding capabilities via a novel multi-motif framework that designs co-occurring motifs with unspecified inter-motif positions and orientations. This makes possible complex protein designs that engage multiple interaction partners and perform multiple functions. On both unconditional and conditional generation, Genie 2 achieves state-of-the-art performance, outperforming all known methods on key design metrics including designability, diversity, and novelty. Genie 2 also solves more motif scaffolding problems than other methods and does so with more unique and varied solutions. Taken together, these advances set a new standard for structure-based protein design. Genie 2 inference and training code, as well as model weights, are freely available at: \url{https://github.com/aqlaboratory/genie2}.
\end{abstract}

\section{Introduction}

The design of proteins with novel structures and functions has emerged as a potent technology in therapeutic \cite{silva2019novo,cao2020novo,shanehsazzadeh2023unlocking} and industrial applications \cite{quijano2021novo,huddy2024blueprinting}. Generative AI has driven recent advances in protein design, most notably diffusion \cite{ho2020denoising,song2020score} and flow matching \cite{lipman2022flow} models, as has the revolution in protein structure prediction sparked by AlphaFold 2 \cite{jumper2021highly}. Proteins are one-dimensional polymers of amino acids ("sequences") that fold into three-dimensional shapes ("structures"). Generative protein models mirror this delineation, with most operating either in the sequence or structural domain. One rationale for sequence-based methods is that sequences are what ultimately get synthesized as functioning biomolecules, while structures require an additional structure-to-sequence map (inverse folding). Sequence-based models include EvoDiff \cite{alamdari2023protein}, a discrete diffusion model that uses order-agnostic autoregressive diffusion with a ByteNet-style \cite{kalchbrenner2016neural} architecture for denoising. EvoDiff is a promising and complementary approach to structure-based design, currently the prevalent paradigm.

Structure-based methods \cite{trippe2022diffusion,wu2024protein,lin2023generating,ingraham2023illuminating,yim2023se,yim2023fast,anand2022protein,fu2024latent,wang2024proteus} focus on modeling structure space and typically employ separate inverse folding models such as ProteinMPNN \cite{dauparas2022robust} to propose plausible sequences given a generated structure. Their key rationale is that structure more closely associates with protein function than sequence. Among them, Genie performs diffusion on backbone atom coordinates and uses an SE(3)-equivariant denoiser to reason over a cloud of reference frames constructed from backbone coordinates. FrameDiff \cite{yim2023se} uses a diffusion process in SE(3) on backbone frames with an AlphaFold-inspired architecture for denoising. FrameFlow \cite{yim2023fast} adopts the general architecture of FrameDiff but uses flow matching instead. Chroma \cite{ingraham2023illuminating} combines a correlated diffusion process that respects statistical properties of natural proteins with an efficient graph neural network. It also includes a separate design network that predicts sequences and side-chain atoms given a generated backbone. More recently, Proteus \cite{wang2024proteus} uses a similar diffusion process and architecture as FrameDiff but introduces graph triangle blocks that combine the expressiveness of triangle attention from AlphaFold 2 with faster runtimes by limiting attention to nearby residues.

The inter-connectedness of sequence and structure suggests that integrating their representations would advance protein design, particularly for conditional tasks that require pre-specified sequence or structural elements. Recent methods reflect this. One approach integrates sequence information as a condition of a structure-based diffusion process, as RFDiffusion \cite{watson2023novo} does when designing proteins with known sequence fragments. Another approach performs diffusion or flow matching in a joint sequence-structure space, as done by MultiFlow \cite{campbell2024generative} when it combines an SE(3) structural flow with a discrete sequence flow. There have also been attempts \cite{costa2023ophiuchus} at jointly encoding sequence and structure in a latent space and diffusing in this space; however, the approach remains nascent.

Whether encoded by sequence or structure, function is what is sought in protein design. Many functions, including interactions with small molecules and other proteins, are governed by few residues, or a \textit{motif}. Achieving prescribed functions can thus often be distilled into designing a protein with a specific motif (\textit{e.g.,} an enzyme active site \cite{wang2022scaffolding} or antigen-binding site \cite{yang2021bottom}), known as \textit{motif scaffolding}. Diffusion models have shown success in this realm: \citet{wu2024practical} developed a sequential Monte Carlo sampler called Twisted Diffusion Sampler and applied it to FrameDiff to scaffold motifs while RFDiffusion and an updated FrameFlow \cite{yim2024improved} were explicitly trained on motif-conditioned tasks. Yet, current models cannot design proteins with multiple independent motifs, as they require inter-motif positions and orientations to be known \textit{a priori}. Proteins often comprise independent functional sites, either as separate domains connected by a flexible linker or as one globular domain, such as an enzyme with multiple substrate binding sites or a scaffolding protein that engages multiple signaling ligands. The ability to design such proteins, which we term \textit{multi-motif scaffolding}, would enable the development of new enzymes \cite{ebrahimi2023engineering}, biosensors \cite{yang2021bottom}, and therapeutics that disrupt or enhance protein-protein interactions \cite{marchand2022computational}. Concurrent with our work, \citet{castro2024accurate} employed an established non-diffusion model, $\text{RF}_\text{joint2}$, to inpaint an immunogen containing three distinct epitopes. This approach appears promising but has yet to be systematically benchmarked.

In this work, we extend Genie to support single- and multi-motif scaffolding. We also improve the core Genie model through architectural modifications and enhancements to its training data and process. The resulting Genie 2 better captures protein structure space. When compared to existing models, Genie 2 sets state-of-the-art results in designability, diversity, and novelty. In addition, Genie 2 surpasses RFDiffusion on motif scaffolding tasks, both in the number of solved problems and the diversity of designs. We also curate a benchmark set comprising 6 multi-motif scaffolding problems from the literature and show that Genie 2 can propose complex designs incorporating multiple functional motifs, a challenge unaddressed by existing protein diffusion models.

\section{Previous Genie Model}

\paragraph{Diffusion with asymmetric protein representations} In contrast to all other SE(3)-equivariant diffusion models for protein generation, which use unified representations for the forward and backward diffusion processes, Genie represents proteins as point clouds of $C_{\alpha}$ atoms in the forward process and as clouds of reference frames in the reverse process. Let $\mathbf{x} = [\mathbf{x}^1, \mathbf{x}^2, \cdots, \mathbf{x}^N]$ be a sequence of $C_\alpha$ coordinates of length $N$. Given a sample $\mathbf{x}_0$ from the unknown protein structure distribution, Genie's forward process gradually adds isotropic Gaussian noise through a cosine variance schedule $\beta = [\beta_1, \beta_2, \cdots, \beta_T]$, where $T$ is the total number of diffusion steps (set to 1,000).
\begin{equation}
    q(\mathbf{x}_t | \mathbf{x}_{t-1}) = \mathcal{N}(\mathbf{x}_t \vert \sqrt{1 - \beta_t} \mathbf{x}_{t-1}, \beta_t \mathbf{I})
\end{equation}
By reparameterization, we have
\begin{equation}
    q(\mathbf{x}_t | \mathbf{x}_{0}) = \mathcal{N}(\mathbf{x}_t \vert \sqrt{\bar{\alpha}_t} \mathbf{x}_0, (1 - \bar{\alpha}_t) \mathbf{I}) \quad \text{where} \quad \bar{\alpha}_t = \prod_{s=1}^t \alpha_s \quad \text{and} \quad \alpha_t = 1 - \beta_t
\end{equation}
Since the isotropic Gaussian noise added at each diffusion step is small, the corresponding reverse process could be approximated with a Gaussian distribution:
\begin{equation}
    p(\mathbf{x}_{t-1} | \mathbf{x}_t) = \mathcal{N}(\mathbf{x}_{t-1} \vert \mathbf{\mu}_\theta(\mathbf{x}_t, t), \mathbf{\Sigma}_\theta(\mathbf{x}_t, t) \mathbf{I})
\end{equation}
where
\begin{equation*}
    \mathbf{\mu}_\theta(\mathbf{x}_t, t) = \frac{1}{\sqrt{\alpha_t}} \left( \mathbf{x}_t - \frac{\beta_t}{\sqrt{1 - \bar{\alpha}_t}} \mathbf{\epsilon}_\theta(F(\mathbf{x}_t), t) \right) \quad \quad \mathbf{\Sigma}_\theta(\mathbf{x}_t, t) = \gamma^2 \cdot \beta_t
\end{equation*}
$F(\cdot)$ is the Frenet-Serret frame construction process based on a sequence of coordinates, and $\gamma \in [0, 1]$ controls the scale of injected noise in the reverse process (analogous to sampling temperature).

\begin{figure}[t]
  \centering
  \includegraphics[width=\textwidth]{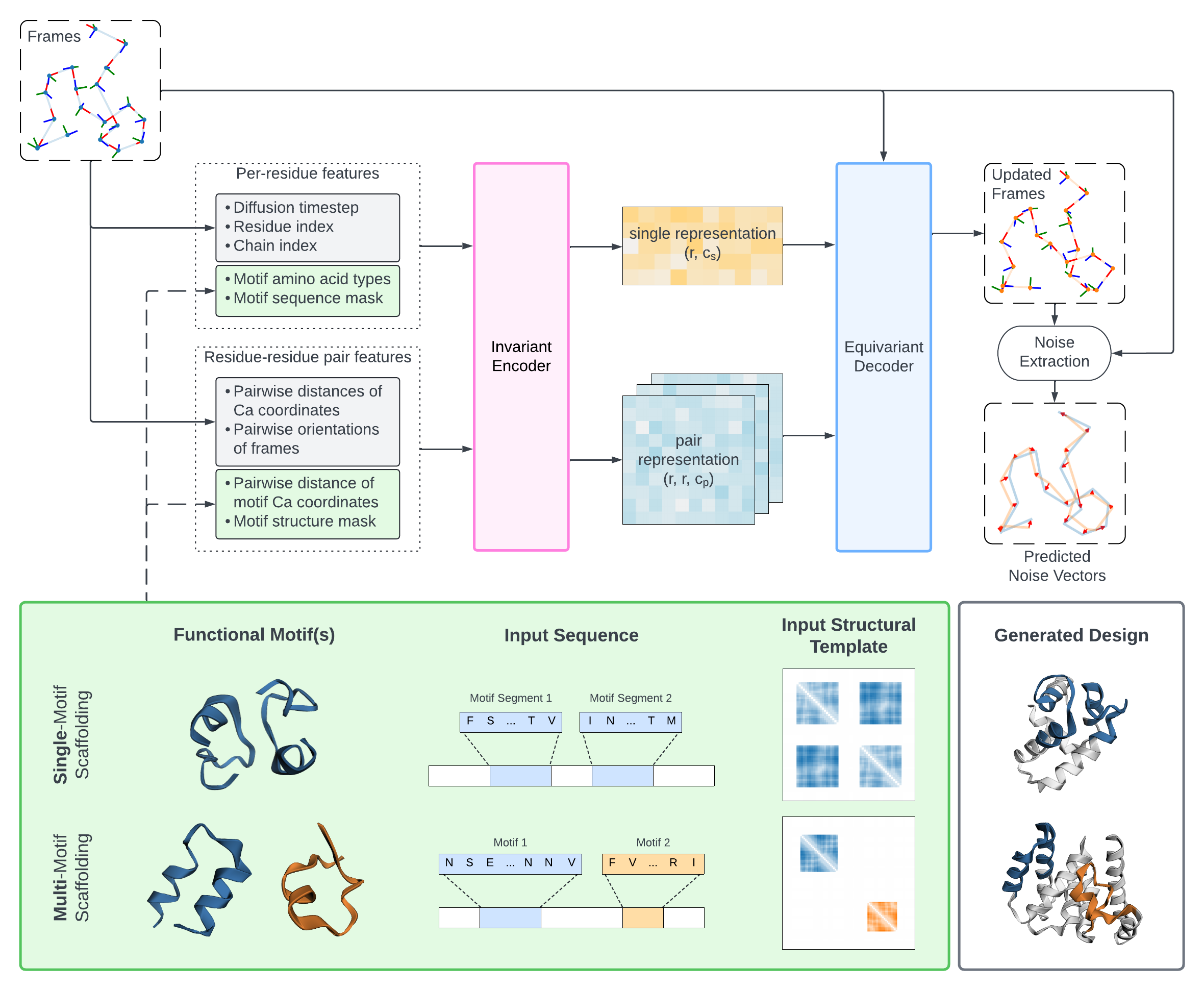}
  \vspace{-.15in}
  \caption{Genie 2 architecture (top), which extends Genie to enable scaffolding on (multiple) motifs. It consists of an SE(3)-invariant encoder that transforms input features into single residue and pair residue-residue representations, and an SE(3)-equivariant decoder that updates frames based on single representations, pair representations, and input reference frames. Example inputs to the model for single- and multi-motif scaffolding problems are shown (bottom-left green box), along with the corresponding generated designs (bottom-right box). In single motif scaffolding (top row), the motif may be contiguous or non-contiguous but all inter-residue positions and orientations are defined. In multi-motif scaffolding (bottom row), inter-motif geometry is left unspecified. For input sequences, white boxes denote masked out regions corresponding to the scaffold.}
\vspace{-.15in}
  \label{fig:architecture}
\end{figure}

\paragraph{SE(3)-equivariant denoiser} The core of Genie is its SE(3)-equivariant denoiser $\epsilon_\theta(F(\mathbf{x}_t), t)$, which reasons over reference frames to predict the noise injected during the forward process. Figure \ref{fig:architecture} summarizes Genie's architecture. The denoiser consists of an SE(3)-invariant encoder, which transforms individual residue and residue-residue pair features into single and pair representations, and an SE(3)-equivariant decoder, which uses Invariant Point Attention \cite{jumper2021highly} to update single representations that are in turn used to update input reference frames. Final noise vectors are computed as the displacement between the translation component of the updated frames and that of the input frames. For more details refer to \cite{lin2023generating}.

\section{Methods}
\label{sec:method}

In this work we extend Genie's architecture and training procedure to enable motif scaffolding. We also substantially improve the core unconditional model through data augmentation and scaling.

\paragraph{Motif representation for conditional generation} Genie's architecture naturally permits integration of conditional sequence and structure information into the diffusion process. We do so by encoding the residues of each motif as one-hot vectors and concatenating these encodings to the single residue features. We encode the structure of each motif using the pairwise distance matrix of its $C_\alpha$ atoms. This representation is SE(3)-invariant as it does not encode the absolute position and orientation of the motif(s), and is unlike the motif conditioning procedures of other methods (\textit{e.g.,} RFDiffusion and FrameFlow), which fix motif coordinates and are thus sensitive to initial placement(s).

Our approach sidesteps a challenge in multi-motif scaffolding, where the design objective leaves the relative positions and orientations of motifs unspecified. By representing motif structures using pairwise distance matrices that specify intra-motif but not inter-motif distances, Genie 2 learns to satisfy the constraints of each motif while generating self-consistent configurations of inter-motif geometries. Figure \ref{fig:architecture} illustrates the types of (multi-)motif templates that can be specified. Note that even in single motif scaffolding, a motif may be non-contiguous by comprising multiple segments. What differentiates single and multi-motif scaffolding is that inter-segment geometric relationships are specified while inter-motif relationships are not. Genie 2's formulation does require specifying sequence length separations between motifs, either by fixing them or sampling from a distribution.

\paragraph{Training} Genie 2 is trained in a purely conditional manner with every training example constituting a (single) motif scaffolding task. Tasks are constructed by first sampling structures from our training dataset to serve as ground truths. A target motif is then constructed for each structure by sampling $N_s$ segments totaling $N_r$ residues, where $N_s \sim \mathcal{U}(1, 4)$, $N_r \sim \mathcal{U}(\lfloor 0.05N \rfloor, \lceil 0.5N \rceil)$, and $N$ is again the length of the protein. The starting positions and lengths of motif segments are randomly chosen subject to the number of motif residues totalling $N_r$. Algorithm 1 describes the task sampling procedure in more detail. We initially experimented with training on varying ratios of conditional and unconditional tasks but found that higher proportions of conditional tasks generally yielded better performance on both types of tasks, and thus switched to purely conditional training. We include an analysis of this behavior in Appendix \ref{app:control_task_ratio}. Due to computational constraints, we limit sequence length to 256 during training; however, Genie 2 is capable of generating proteins longer than 256 residues. In addition, we do not train on multi-motif scaffolding as our input representation permits under-specification of geometric relationships as an inference-time choice. Genie 2's performance on multi-motif scaffolding thus represents out-of-distribution generative generalization.

\begin{algorithm}
\caption{Motif construction for conditional training task}\label{alg:cap}
\begin{algorithmic} \onehalfspacing
\Require Sampled structure $\mathbf{x}$, a sequence of $C_\alpha$ coordinates of length $N$
\State $N_s \sim \mathcal{U}(1, 4)$ \Comment{Number of segments in the motif}
\State $N_r \sim \mathcal{U}(\lfloor 0.05N \rfloor, \lceil 0.5N \rceil)$ \Comment{Number of residues in the motif}
\State $B \leftarrow [0, b_1, b_2, \cdots, b_{N_s - 1}, N_r]$ \quad where $b_1, b_2, \cdots, b_{N_s - 1}$ 
\State \quad \quad are randomly sampled from $\{1, 2, \cdots, N_r - 1\}$
\State \quad \quad without replacement and sorted in ascending order.
\State $L \leftarrow [l_1, l_2, \cdots, l_{N_s}]$ \quad where $L_i = B_i - B_{i-1}$ \Comment{Split motif residues into segments}
\State $\mathbf{M} = \text{Flatten}(\text{Permute}([S_1, S_2, \cdots, S_{N - N_r}, M_1, M_2, \cdots, M_{N_s}]))$ \quad where
\State \quad \quad $S_i = [0]$ for $i \in [1, N - N_r]$ \Comment{Represents a scaffold residue}
\State \quad \quad $M_j = [1, 1, \cdots, 1]$ where $|M_j| = l_j$ for $j \in [1, N_s]$ \quad \quad \Comment{Represents a motif segment}
\Return $\mathbf{M}$ \quad where for $i \in [1, N]$ \Comment{Represents a motif sequence mask}
\State \quad \quad $\mathbf{M}[i] = 1$ indicates that residue $i$ is a motif residue
\State \quad \quad $\mathbf{M}[i] = 0$ indicates that residue $i$ is a scaffold residue
\end{algorithmic}
\end{algorithm}

\paragraph{Data augmentation} Diffusion models require large datasets to robustly capture complex distributions. Generative protein models have thus far relied on training on experimentally determined protein structures from the Protein Data Bank (PDB) \cite{berman2002protein,burley2023rcsb}. Despite the enormous experimental efforts that have gone into assembling the PDB, its size remains limited to \~{}20,000 proteins of relevant lengths. With the development of highly accurate protein structure prediction, we hypothesized that augmenting Genie training with confidently predicted protein structures could boost its performance by expanding the space of observed folds beyond those present in the PDB. Consequently, we train Genie 2 using the AlphaFold database (AFDB) \cite{varadi2022alphafold}, which consists of approximately 214M AlphaFold 2 predictions spanning nearly the entirety of UniProt  \cite{uniprot2023uniprot}. As AFDB is highly structurally redundant, we use a subsampled version \cite{barrio2023clustering} that applies FoldSeek \cite{van2024fast} to cluster entries based on structural similarity. We start with all cluster representatives from the FoldSeek-clustered database and then filter them using a pLDDT threshold of >80, to enrich for highly confident predictions, and a maximum sequence length of 256. This results in 588,570 structures. To our knowledge, Genie 2 is the first protein diffusion model to train on AFDB.

\paragraph{Loss function} We minimize the loss function below, which computes the mean squared error between predicted and ground truth noise:
\begin{align}
    L(\theta) &= \mathbb{E}_{t, x_0, \epsilon} \left[ \frac{1}{N} \sum_{i=1}^{N} \left\Vert  \epsilon_t^i  - \epsilon_\theta^i(F(x_t), t) \right\Vert^2 \right] \\
    &= \mathbb{E}_{t, x_0, \epsilon} \left[ \frac{1}{\vert \mathcal{M} \vert + \vert \mathcal{S} \vert} \left(\sum_{i \in \mathcal{M}} \left\Vert  \epsilon_t^i  - \epsilon_\theta^i(F(x_t), t) \right\Vert^2 + \sum_{i \in \mathcal{S}} \left\Vert \epsilon_t^i  - \epsilon_\theta^i(F(x_t), t) \right\Vert^2 \right) \right]
\end{align}
where $\mathcal{M}$ and $\mathcal{S}$ are the set of motif and scaffold residue indices, respectively. Under this construction, motifs are enforced as a soft constraint, ensuring that the model is responsive to motif specifications while also designing the protein as a whole.

\section{Unconditional Protein Generation}

To systematically assess Genie 2 and competing methods on unconditional protein generation, we conduct two sets of analyses. First, we assess methods without accounting for length while restricting the longest designed protein to 256 residues (Section \ref{uncond_indist}). This reflects Genie 2's in-distribution generative power since it is trained on proteins up to 256 residues long. Second, we assess methods in a length-specific manner up to 500 residues (Section \ref{uncond_length}) to quantify Genie 2's out-of-distribution generative capabilities. In both analyses, we rely on the evaluation metrics described in Section \ref{uncond_eval_metric}.

We compare Genie 2 to Chroma, FrameFlow and RFDiffusion. The latter is widely perceived as the current state-of-the-art protein design model and has been extensively validated. A more recent model, Proteus, asserts some gains on designability over RFDiffusion at the cost of lower diversity. Unfortunately, the code for Proteus is not publicly available, precluding direct comparison. Nonetheless, based on Proteus' reported metrics, we believe that the comparison with RFDiffusion is sufficient to establish Genie 2 as the new state-of-the-art model. We note that while Chroma contains a built-in sequence design network, we find it to underperform ProteinMPNN and so exclude it; instead we adopt the same evaluation pipeline across all methods.

\subsection{Evaluation metrics}
\label{uncond_eval_metric}

\paragraph{Designability} A structure that can be plausibly realized by some protein sequence is one that is designable. To determine if a structure is designable we employ a commonly used pipeline \cite{trippe2022diffusion} that computes \textit{in silico} self-consistency between generated and predicted structures. First, a generated structure is fed into an inverse folding model (ProteinMPNN \cite{dauparas2022robust}) to produce 8 plausible sequences for the design. Next, structures of proposed sequences are predicted (using ESMFold \cite{lin2022language}) and the consistency of predicted structures with respect to the original generated structure is assessed using a structure similarity metric (TM-score \cite{zhang2004scoring,xu2010significant}). Using this pipeline, we consider a generated structure to be designable if it is within 2Å RMSD of the most similar predicted structure ($\text{scRMSD} \leq 2$) and the structure is confidently predicted (mean $\text{pLDDT} \geq 70$). Over a set, "designability" quantifies the fraction of designable structures within it. We note that designability alone can be misleading because it does not account for structural diversity--for example, a model that has mode-collapsed into a single designable structure achieves perfect designability.

\paragraph{Diversity} Complementary to designability is the (structural) diversity of a generated protein set. To quantify diversity we start by hierarchically clustering (with single linkage) the set of \textit{designable} generated structures. We exclude non-designable structures as we do not expect them to be realizable and including them would thus inflate diversity. As sequence lengths vary within a set of designable structures, we utilize TMAlign \cite{zhang2005tm} to compute the pairwise similarities of all structures and use a TM-score threshold of 0.6 as cutoff. This implies that any pair of structures across clusters would have a TM score of at most 0.6. We then compute "diversity" as the fraction of distinct designable clusters within a set of generated structures. As the diversity metric already enforces designability of generated clusters, we find that it better reflects the generative capabilities of a model than designability. Note that diversity depends on the number of samples generated, and tends to 0 as sample size increases. In all our experiments we use a fixed sample size to enable even comparisons.

\paragraph{F1 score} Following \citet{lin2023generating}, we compute the harmonic mean between designability ($p_{\text{structures}}$) and diversity ($p_{\text{clusters}}$) as follows:
\begin{equation}
    F_\beta = (1 + \beta^2) \cdot \frac{p_{\text{structures}} \cdot p_{\text{clusters}}}{\beta^2 \cdot p_{\text{structures}} + p_{\text{clusters}}}
\end{equation}
where $\beta \in \mathbb{R}^+$ controls the relative weighting of designability and diversity. We set $\beta = 1$ and report the metric as F1 score.

\paragraph{Novelty} Beyond designability and diversity, we also quantify the novelty of generated structures with respect to reference  datasets and, by extension, the known structural universe. To compute the novelty of a generated structure we again employ TM-score as our structure similarity metric and use TMAlign to compute the TM scores between a generated structure and all structures in a reference dataset. We consider a generated structure to be novel if it is designable and its TM-score to any reference structure is at most 0.5. Similar to our diversity calculations, we apply hierarchical clustering (with single linkage and a TM-score threshold of 0.6) to the set of novel structures and define "novelty" to be the fraction of distinct novel clusters within a set of generated structures. We measure novelty with respect to both the PDB and Foldseek-clustered AFDB datasets (the latter being our training dataset) and term these measures "PDB novelty" and "AFDB novelty", respectively.

\subsection{In-distribution performance analysis}
\label{uncond_indist}

\begin{table*}
\caption{Unconditional generative performance of structure-based diffusion models.}
\label{tbl:unconditional}
\vskip -0.15in
\begin{center}
\begin{sc}
\begin{small}
\begin{tabular}{lcccccccr}
\toprule
Method & Designability & Diversity & $F_1$ & PDB Novelty & AFDB Novelty \\
\midrule
Chroma & 0.70 & 0.51 & 0.59 & 0.13 & 0.04 \\
RFDiffusion & \textbf{0.96} & 0.63 & 0.76 & 0.26 & 0.14 \\
Genie 2 & \textbf{0.96} & \textbf{0.91} & \textbf{0.93} & \textbf{0.41} & \textbf{0.21} \\
\bottomrule
\end{tabular}
\end{small}
\end{sc}
\end{center}
\vskip -0.2in
\end{table*}

\begin{figure}[h!]
  \centering
  \includegraphics[width=\textwidth]{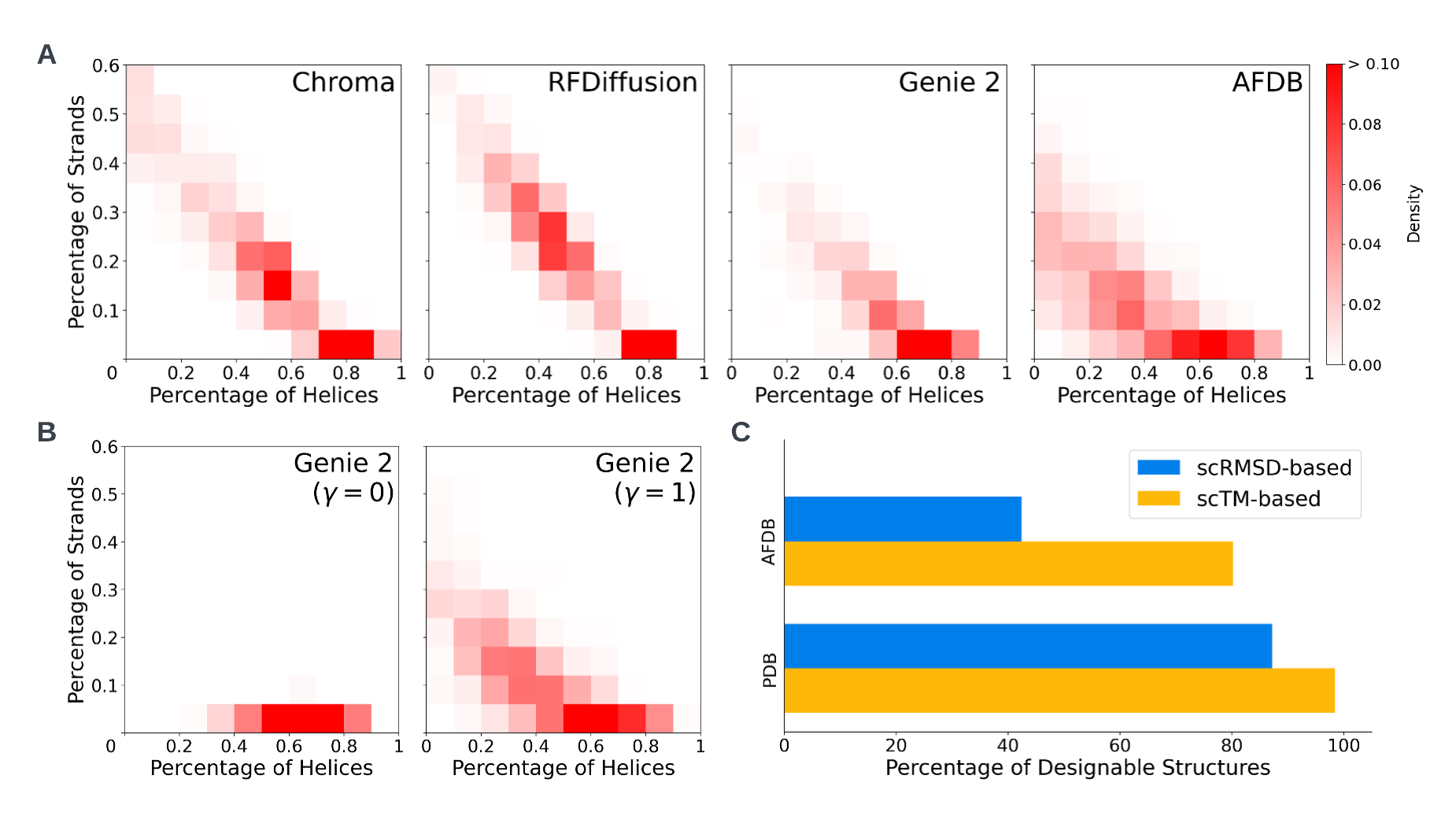}
  \vspace{-.2in}
  \caption{Visualizations of in-distribution performance on unconditional generation. (A) Secondary structure distributions of proteins generated by Chroma, RFDiffusion and Genie 2. For reference, we also include the secondary structure distribution of 1,000 structures randomly drawn from AFDB (far right). (B) Secondary structure distributions of proteins generated by Genie 2 when sampling noise scale ($\gamma$ in equation (3)) is set to 0 (left) and 1 (right). (C) Self-consistency results on 1,000 randomly chosen structures from the PDB and clustered AFDB datasets.}
  \label{fig:uncond_result_secondary}
  \vskip -0.1in
\end{figure}

We assess Genie 2, Chroma, and RFDiffusion by generating 5 structures of every length ranging from 50 to 256 residues (1,035 structures in total). We omit FrameFlow here since it is trained using a maximum sequence length of 128, but include direct comparisons with FrameFlow in Section \ref{uncond_length}. Table \ref{tbl:unconditional} summarizes the performance of all methods on our key metrics. Relative to RFDiffusion and Chroma, Genie 2 achieves comparable designability and much higher diversity and novelty. This suggests that as a core unconditional model, Genie 2 best captures foldable protein structure space, and may thus serve as a superior engine for downstream sampling-based protein design tasks \cite{wu2024practical,didi2023framework}.

In Figure \ref{fig:uncond_result_secondary}A, we visualize the secondary structure distribution of generated proteins. While all methods yield a wide range of secondary structure elements, the resulting distributions are biased (relative to AFDB), with beta strand-containing structures (top left of distribution) and loop elements (bottom left) being generally underrepresented. There are multiple possible reasons for this bias. First, the high frequency of helices in the training dataset leads to models that favor generating helical structures. Second, alpha helices are likely easier to generate than beta sheets as they involve largely local interactions while sheets may involve long-range interactions. Third, we assess Genie 2 using low temperature sampling as it yields better results, but this may shift the model from its learned distribution. We test this hypothesis by visualizing the distribution of secondary structures generated by Genie 2 under a normal temperature ($\gamma = 1$) in Figure \ref{fig:uncond_result_secondary}B. We observe that the resulting distribution is in fact consistent with that of the clustered AFDB dataset.

This raises the question of whether low temperature sampling is necessary or, alternatively, why it helps improve Genie 2's performance. To investigate this we ran our self-consistency pipeline on 1,000 randomly chosen structures from the clustered AFDB dataset. We found that only 42.4\% of these structures are designable. When our designability criteria is relaxed to $\text{scTM} \geq 0.5$ and $\text{pLDDT} \geq 70$, the percentage of designable structures increases to 80.2\% (Figure \ref{fig:uncond_result_secondary}C). For comparison, among PDB structures 87.2\% are considered designable by our original criteria. This suggests that while AF-predicted structures are globally reliable (with $\text{scTM} \geq 0.5$), local atomic details remain much less accurate. Since we use AFDB for training, this might explain why designability is low at normal temperature sampling, as lower temperatures appear to bias Genie 2 towards higher fidelity. We note that ProteinMPNN was exclusively trained on the PDB and thus we cannot rule out the possibility that the apparent discrepancy in designability between the PDB and AFDB is due to a bias in ProteinMPNN towards the PDB.

\subsection{Length-based performance analysis}
\label{uncond_length}

\begin{figure}
  \centering
  \includegraphics[width=\textwidth]{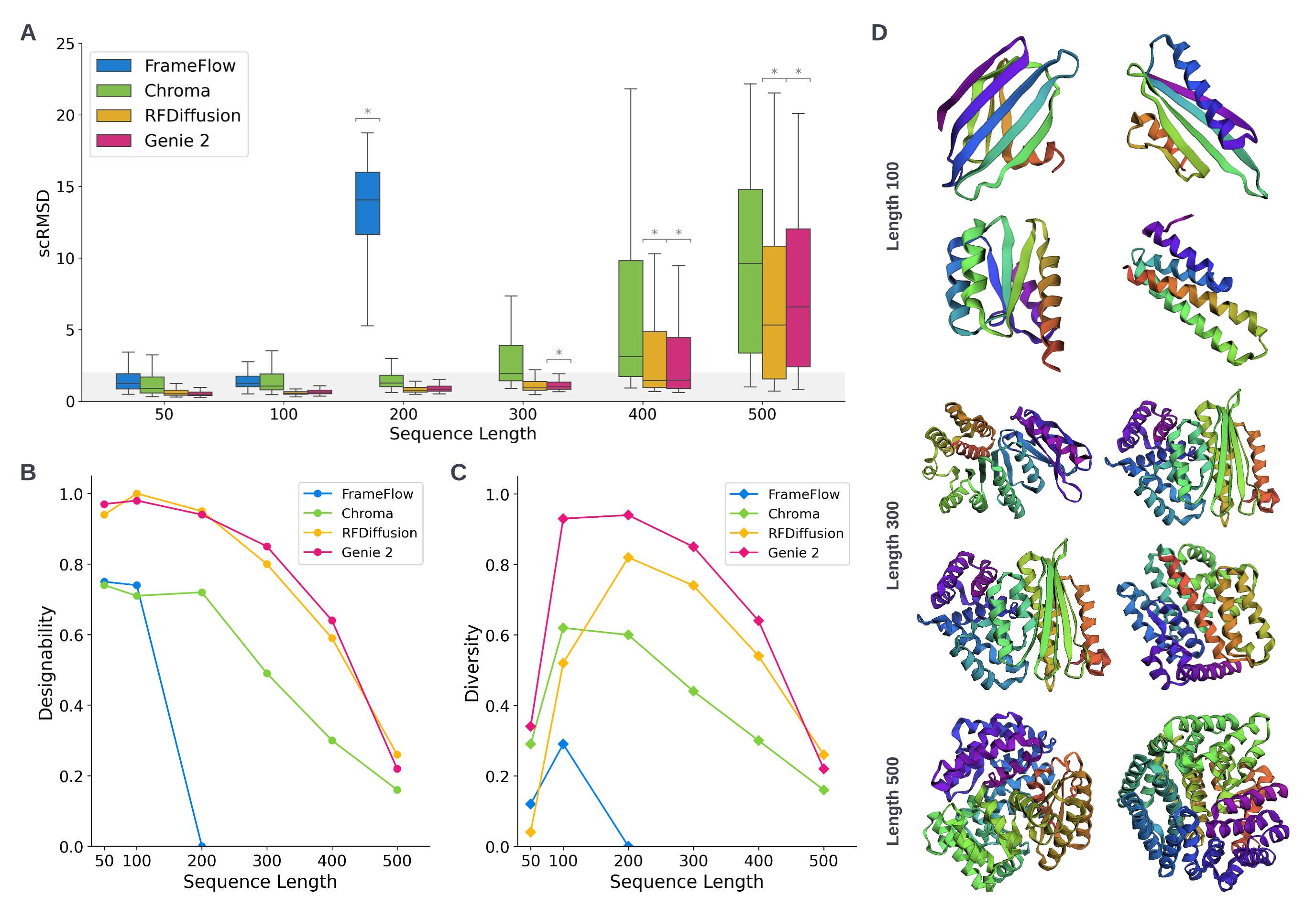}
  \vspace{-.2in}
  \caption{Assessment of methods by sequence length. For each method/sequence length combination, we generate 100 structures. (A) Box-and-whisker plots of scRMSDs between generated structures and their most similar ESMFold-predicted structures. Asterisks (*) indicate that sequence lengths exceed the maximum seen during training. (B-C) Plots of designability (B) and diversity (C) as a function of sequence length. (D) Example structures generated by Genie 2.}
  \label{fig:uncondresults_length}
  \vskip -0.1in
\end{figure}

We next assess generative performance in a length-dependent manner. For a subset of sequence lengths ranging from 50 to 500 residues, we generate 100 structures and assess them using our design metrics. Figure \ref{fig:uncondresults_length}A shows the scRMSD distribution across sequence lengths while Figures \ref{fig:uncondresults_length}B and \ref{fig:uncondresults_length}C plot designability and diversity as a function of sequence length, respectively. At nearly all assessed lengths, Genie 2 has comparable designability to RFDiffusion but higher diversity. For short proteins (<200 residues), Genie 2 exhibits considerably higher diversity (doubling that of RFDiffusion at 100 residues), which is noteworthy as shorter lengths constitute smaller design spaces.

Generative models generally struggle with creating larger proteins due to their increased complexity. As sequence length increases, designability decreases and in turn so does diversity, likely because diversity depends on the number of designable proteins. Larger protein lengths should in principle permit greater diversity but they are harder to generate. Nonetheless, despite having been trained on monomers of at most 256 residues, Genie 2 can generate 500-residue structures with comparable or better performance than competing methods. For reference, RFDiffusion uses a crop size of 384 during training while Chroma trains on even larger proteins (>500 residues) owing to its efficient graph neural network. Figure \ref{fig:uncondresults_length}D shows examples of Genie 2 designed structures of varying lengths.

\section{Motif Scaffolding}

In this section we assess Genie 2 on single motif scaffolding and compare it to RFDiffusion on the same set of design tasks. A more recent method, FrameFlow, does assert superiority over RFDiffusion on single motif scaffolding, but its scaffolding-capable code is unfortunately unavailable, precluding direct comparison. We additionally assess Genie 2 on multi-motif scaffolding using a suite of 6 multi-motif tasks that we curated for this assessment.

\begin{figure}
  \centering
  \includegraphics[width=\textwidth]{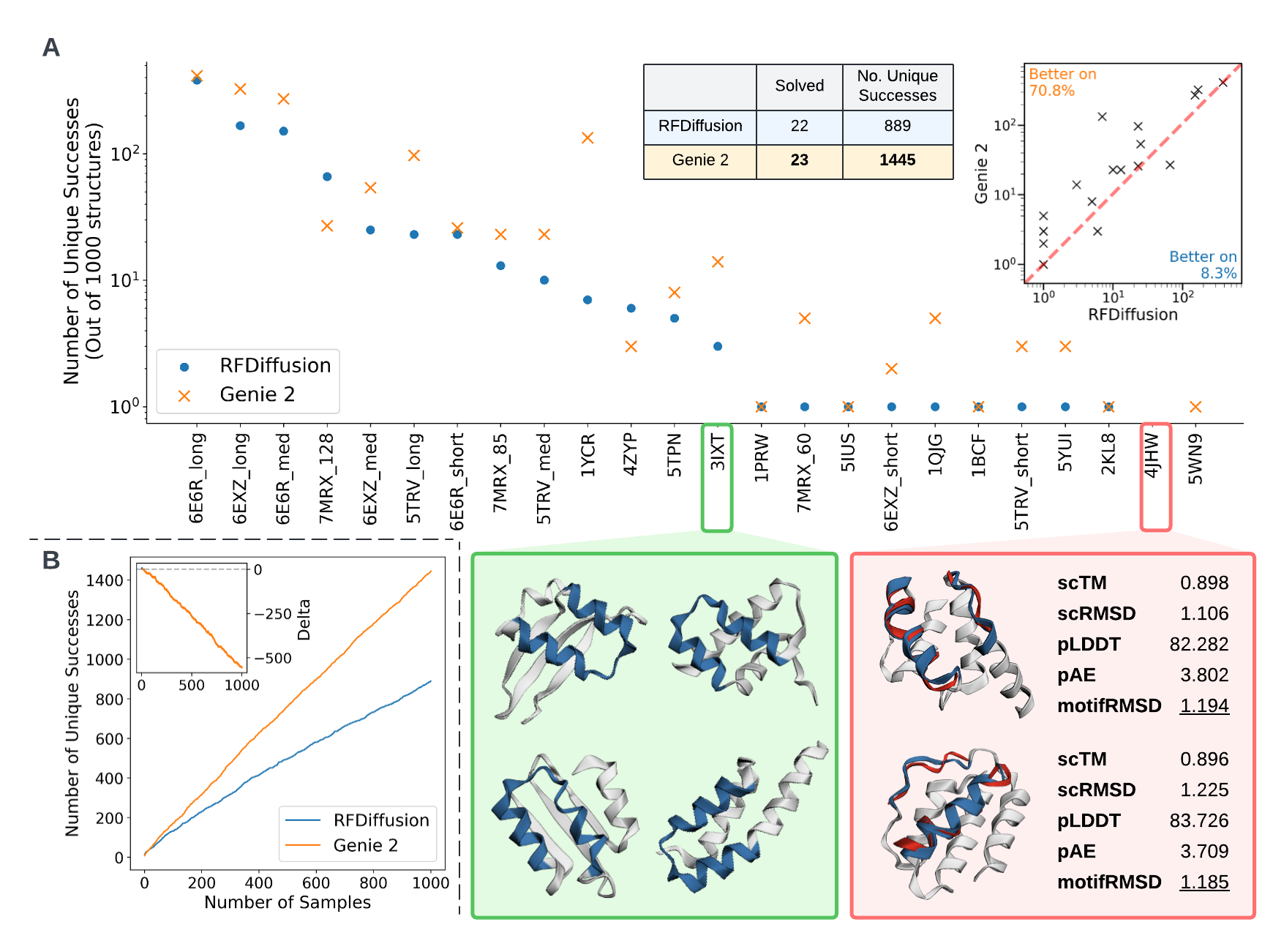}
  \vspace{-.3in}
  \caption{Comparison of Genie 2 and RFDiffusion on single-motif scaffolding. (A) Performance of Genie 2 and RFDiffusion across 24 single-motif scaffolding tasks. Inset (top right) shows a scatter plot of the (unique) success rate of Genie 2 vs. RFDiffusion; each point represents a scaffolding task. Summary statistics are shown in table (left). Example designs are shown (bottom) for successful task 3IXT (green) as well as failed task 4JHW (red). Scaffolds (white), motifs (blue), and unsatisfied sought motifs (red) are overlaid. (B) Plot of number of unique successes as a function of sample size.}
  \label{fig:scaffoldresults}
   \vskip -.1in
\end{figure}

\subsection{Evaluation metrics}

Motif scaffolding problems consist of sequence and structure constraints on motif(s) plus length (min/max) constraints on scaffolds and the overall protein. To solve a motif scaffolding problem, we first sample a constraint-satisfying length for each scaffold segment while ensuring that total protein length is also within specifications. This information, together with the sequence and structure of motif(s), is passed as conditions to Genie 2. We quantify success using the criteria of RFDiffusion, which requires that generated structures achieve $\text{scRMSD} \leq 2$Å, $\text{pLDDT} \geq 70$, and $\text{pAE} \leq 5$ to be considered designable, and for designed motif(s) to have backbone $\text{RMSD} \leq 1$Å with respect to each motif to be considered constraint satisfying.

While most previous studies, including RFDiffusion, use success rate as the evaluation metric, we find that this tends to inflate performance, as it is possible to achieve high success rates by repeatedly generating only one or a few successful designs, \textit{i.e.,} while suffering from mode collapse. Instead, and similar to \cite{yim2024improved}, we cluster successful designs based on structure similarity and report the number of unique successes. This approach better balances designability with diversity when assessing motif scaffolding performance. We use hierarchical clustering with single linkage and a TM-score threshold of 0.6. For each motif scaffolding problem, we sample 1,000 structures.

\begin{figure}
  \centering
  \includegraphics[width=\textwidth]{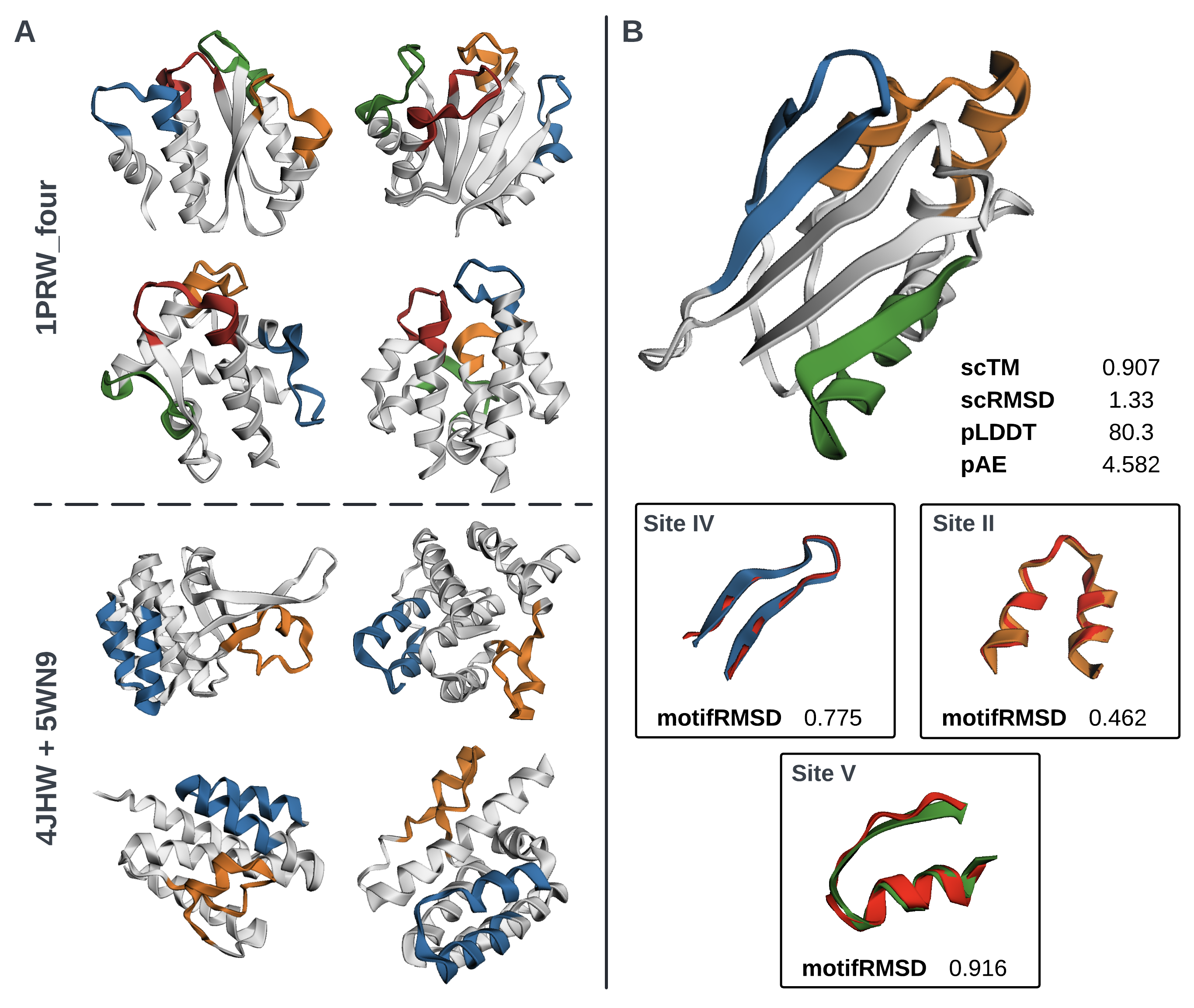}
  \vspace{-.2in}
  \caption{Performance of Genie 2 on multi-motif scaffolding tasks. (A) Successful designs for task 1PRW\_four (scaffolding with four $\text{Ca}^{2+}$ ion binding sites) and 4JHW+5WN9 (scaffolding with RSV-F site II epitope and RSV-G 2D10 epitope). Scaffolds are in grey and distinct motifs are colored differently. (B) (Top) Successful design for multi-epitope immunogen. (Bottom) Individual epitope designs superposed over target structures (red). }
  \label{fig:multiscaffoldresults}
   \vspace{-.15in}
\end{figure}

\subsection{Single-motif scaffolding} 
\label{sec:single_motif}

For evaluation, we use a previously published motif scaffolding benchmark \cite{watson2023novo} comprising 25 tasks curated from six recent publications. We exclude one task, 6VW1, as its motif consists of segments from multiple protein chains, a requirement not supported by Genie 2. Figure \ref{fig:scaffoldresults}A summarizes the performance of Genie 2 and RFDiffusion across the remaining 24 tasks. For 22 tasks, Genie 2 yields a similar or larger number of unique designs than RFDiffusion. Genie 2 also solves task 5WN9 that involves scaffolding the RSV G-protein 2D10 site \cite{chene2003inhibiting}, while RFDiffusion does not. We speculate that Genie 2 solves this task because it is capable of generating more diverse designs. We show examples of successful designs in Figure \ref{fig:scaffoldresults}A with more in Appendix \ref{app:single_result_examples}. In addition, we observe that the performance gap (in terms of number of unique successes) between Genie 2 and RFDiffusion widens as sample size increases (Figure \ref{fig:scaffoldresults}B), suggesting that Genie 2 is capturing a larger and more diverse structure space than RFDiffusion. Genie 2 does fail on one problem, 4JHW, that RFDiffusion also fails on, which involves scaffolding the RSV F-protein site-0. To better understand this failure case, we visualize the two closest designs (red box in Figure \ref{fig:scaffoldresults}) and observe that while Genie 2 yields designable structures it does not satisfy the motif constraints.

\subsection{Multi-motif scaffolding}
\label{sec:multimotif}

To assess the multi-motif scaffolding capabilities of Genie 2, we curate from the literature a set of 6 scaffolding tasks that each require multiple motifs: designing an immunogen with two epitopes \cite{yang2021bottom}, scaffolding two $\text{Ca}^{2+}$ binding sites (four EF hand motifs) \cite{wang2022scaffolding,fallon2003closed}, scaffolding two binding motifs to PD-1 protein \cite{bryan2021computational}, scaffolding $\text{Cl}^{-}$ and $\text{Ni}^{2+}$ binding sites \cite{agnew2011correlation,chalkley2022novo}, and designing a binder of two different proteins, IL-2 receptor $\beta\gamma_c$ heterodimer (IL-2R$\beta\gamma_c$) and IL-2R$\alpha$ \cite{ren2022interleukin,silva2019novo}. More details on this set of tasks are included in Appendix \ref{app:multimotif_benchmark}. The set is meant to reflect the breadth of potential protein design tasks, including immunogen, binder, and enyzme design.

Genie 2 solves 4 of the 6 tasks. Figure \ref{fig:multiscaffoldresults}A shows successful designs of task 1PRW\_four (scaffolding with four $\text{Ca}^{2+}$ ion binding sites) and 4JHW+5WN9 (scaffolding with RSV-F site II epitope and RSV-G 2D10 epitope). More results are included in Appendix \ref{app:multimotif}. In addition to our benchmark set, we apply Genie 2 to a multi-motif task proposed in a concurrent preprint \cite{castro2024accurate}. This task scaffolds an immunogen containing three unique epitopes from the respiratory syncytial virus (RSV) fusion protein. Genie 2 solves the task with only 1,000 samples. Figure \ref{fig:multiscaffoldresults}B shows an example design.

\section{Limitations and Future Work}

\label{sec:limitations}
Genie 2 achieves state-of-the-art performance on both unconditional generation and motif scaffolding. Yet, its sampling time is longer than that of other methods, requiring 1,000 denoising iterations vs. 100 (FrameFlow), 500 (Chroma), and 50 (RFDiffusion). Appendix \ref{app:sample_time} provides a summary of sampling times across sequence lengths. One future direction for Genie 2 is to improve its sampling efficiency in both unconditional protein generation and motif scaffolding. Genie 2 also employs triangular multiplicative update layers, introduced in AlphaFold 2. These layers are computationally expensive with $O(N^3)$ scaling, thus disproportionately affecting larger design tasks. A second future direction is thus to reduce the time and space complexity of the Genie 2 architecture, to enable generation of and training on larger proteins.

\clearpage

\bibliography{reference}
\bibliographystyle{plainnat}

\clearpage 
\appendix

\section{Additional Details on Genie 2}
\label{app:detail}

\subsection{Hyperparameter choices}

In Table 2 we detail the key hyperparameters of the Genie 2 architecture and highlight differences from the original Genie model. For Genie 2, we increase input embedding and single representation dimensions as we found this improves performance without substantially impacting training speed. Due to the increase in model complexity, Genie 2 consists of 15.7M trainable parameters, $\sim$4x the original Genie architecture. However, Genie 2 remains four times smaller than RFDiffusion, which has 59.8M trainable parameters.

\begin{table}[h]
    \centering
    \caption{Key hyperparamters of Genie and Genie 2. Updated values are indicated in \textbf{bold}.}
    \label{tbl:appHyperParams}
    \begin{tabular}{llcc}
    \toprule
        \multicolumn{2}{l}{Hyperparameter} & Genie & Genie 2 \\
    \midrule
        \multicolumn{2}{l}{Number of parameters} & 4.1M & \textbf{15.7M} \\
    \midrule
        \multirow{3}{*}{Input embedding dimension} & Residue index & 128 & \textbf{256} \\
        & Chain index & - & \textbf{64} \\
        & Diffusion timestep & 128 & \textbf{512} \\ 
    \midrule
        \multirow{2}{*}{Representation dimension} & Single representation & 128 & \textbf{384} \\
        & Pair representation & 128 & 128 \\
    \midrule
        \multirow{1}{*}{SE(3)-equivariant decoder} & Number of IPA layers & 5 & \textbf{8} \\
    \bottomrule
    \end{tabular}
    \vskip -.1in
\end{table}

\subsection{Training}
\label{app:training_details}

For training, we use the Adam \cite{kingma2014adam} optimizer with a constant learning rate of $10^{-4}$. We train Genie 2 using data parallelism on 8 Nvidia A100 GPUs with an effective batch size of 48. We train the model for 40 epochs ($\sim$5 days) for a total of $\sim$960 GPU hours. In comparison, RFDiffusion is initialized with pretrained weights from RoseTTAFold \cite{baek2021accurate}, whose training requires 64 Nvidia V100 GPUs for 4 weeks. Training of RFDiffusion takes 3 days on 8 Nvidia A100 GPUs. Hence, Genie 2 requires much less computational resources to train than RFDiffusion.

\subsection{Sampling}

To improve designability, we adjusted the sampling noise scale ($\gamma$ in Equation (3)) to trade diversity for designability. We set $\gamma = 0.6$ and $\gamma = 0.4$ for unconditional generation and motif scaffolding, respectively, as these settings provided the best results. Moreover, for motif scaffolding, we use the checkpoint at epoch 30 since it gives slightly better performance.

\clearpage

\subsection{Effect of varying conditional task ratio}
\label{app:control_task_ratio}

We experimented with varying the frequency of conditional vs. unconditional tasks during training (0.0, 0.2, 0.5, 0.8, and 1.0). Due to computational constraints, we trained models only up to 10 epochs. Although models do not fully converge, we believe the trends are still indicative of final performance. To evaluate models, we follow the same procedure from Section \ref{uncond_indist} when assessing unconditional generation performance: for each model, we generate 5 samples per sequence length ranging from 50 to 256 residues. For single-motif scaffolding evaluation, we use the pipeline described in Section \ref{sec:single_motif}, but sample only 100 designs per motif scaffolding problem to conserve computational costs.

Table \ref{tbl:byTaskRatio} summarizes the performance of these models on both unconditional protein generation and single motif scaffolding. As the conditional task ratio increases, motif scaffolding performance generally improves, with the best performance achieved when the conditional task ratio equals 1. Surprisingly, unconditional generation performance fluctuates but is ultimately also maximized when conditional tasks are exclusively sampled. As a result we use a conditional task ratio of 1 during all training runs.

\begin{table*}[h!]
\caption{Unconditional generation and motif scaffolding performance by conditional task ratio. \textsc{Successes} denote total number of unique successes across all problems.}

\label{tbl:byTaskRatio}
\begin{center}
\begin{sc}
\begin{tabular}{ccccccr}
\toprule
\multirow{2}{*}{Ratio} & \multicolumn{3}{c}{Unconditional Generation} & \multicolumn{2}{c}{Motif Scaffolding} \\
\cmidrule(lr){2-4}
\cmidrule(lr){5-6}
& Designability & Diversity & $F_1$ & Solved & Successes \\
\midrule
0.0 & 0.858 & 0.771 & 0.812 & 1 & 6 \\
0.2 & 0.892 & 0.752 & 0.816 & 14 & 101 \\
0.5 & 0.740 & 0.649 & 0.692 & 13 & 98 \\
0.8 & 0.865 & 0.783 & 0.822 & 18 & 179 \\
\textbf{1.0} & \textbf{0.898} & \textbf{0.802} & \textbf{0.847} & \textbf{19} & \textbf{202} \\
\bottomrule
\end{tabular}
\end{sc}
\end{center}
\end{table*}

\clearpage

\section{Multi-Motif Scaffolding Benchmark}

\label{app:multimotif_benchmark}

In Table \ref{tab:multimotif_config}, we provide detailed configurations for each multi-motif scaffolding task. We name each problem using the names of PDB structures that contain the motif(s) used in the problem. Additional postfixes are added to distinguish between problems whose motifs come from the same PDB structure. In the third column ("configuration"), we provide a detailed input specification for each multi-motif scaffolding problem. Each bolded part denotes a motif segment, including its location in the PDB structure. For example, "5WN9/A170-189\{2\}" in problem 4JHW+5WN9 indicates that the motif segment comes from residue 170 - 189 of Chain A in the protein 5WN9, and "2" (in curly bracket) indicates that this motif segment belongs to the second motif. Each non-bolded part denotes a scaffold segment with minimum and maximum lengths specified. At sampling time, each scaffold length is sampled within this range. For example, "10-40" in problem 4JHW+5WN9 indicates that the scaffold has a length between 10 and 40 (inclusive). The last column ("Total length") specifies the minimum and maximum length requirements for the whole sequence.

\begin{table}[h]
    \caption{\textbf{The benchmark set of multi-motif scaffolding problems}.}
    \label{tab:multimotif_config}
    \centering
    \setlength{\tabcolsep}{3pt}
    \small
    \begin{tabular}{llp{61mm}@{\hspace{10pt}}l}
        \toprule
         Name & Description & Configuration & Length \\
         \midrule
         4JHW+5WN9 \cite{yang2021bottom} & Two epitopes & 
         10-40, \textbf{4JHW/F254-278\{1\}}, 20-50, \textbf{5WN9/A170-189\{2\}}, 10-40 & 85-175 \\
         \midrule
         1PRW\_two \cite{wang2022scaffolding,fallon2003closed} & Two 4-helix bundles & 5-20, \textbf{1PRW/A16-35\{1\}}, 10-25, \textbf{1PRW/A52-71\{1\}}, 10-30, \textbf{1PRW/A89-108\{2\}}, 10-25, \textbf{1PRW/A125-144\{2\}}, 5-20 & 120-200 \\
        \midrule
        1PRW\_four \cite{wang2022scaffolding,fallon2003closed} & Four EF-hands & 5-20, \textbf{1PRW/A21-32\{1\}}, 10-25, \textbf{1PRW/A57-68\{2\}}, 10-25, \textbf{1PRW/A94-105\{3\}}, 10-25, \textbf{1PRW/A125-144\{4\}}, 5-20 & 88-163 \\
         \midrule
        3BIK+3BP5 \cite{bryan2021computational} & Two PD-1 binding motifs & 5-15, \textbf{3BIK/A121-125\{1\}}, 10-20, \textbf{3BP5/B110-114\{2\}}, 5-15 & 30-60 \\
        \midrule
        3NTN \cite{agnew2011correlation,chalkley2022novo} & Two 3-helix bundles & \textbf{\textbf{3NTN/A342-348\{1\}}, 10-10, 3NTN/A367-372\{2\}}, 10-20, \textbf{3NTN/B372-367\{2\}}, 10-10, \textbf{3NTN/B348-342\{1\}}, 10-20, \textbf{3NTN/C342-348\{1\}}, 10-10, \textbf{3NTN/C367-372\{2\}} & 89-109 \\
        \midrule
        2B5I \cite{ren2022interleukin,silva2019novo} & Two binding sites & 5-15, \textbf{2B5I/A11-23\{2\}}, 10-20, \textbf{2B5I/A35-45\{1\}}, 10-20, \textbf{2B5I/A61-72\{1\}}, 5-15, \textbf{2B5I/A81-95\{2\}}, 20-30, \textbf{2B5I/A119-133\{2\}} & 116-166 \\
        \bottomrule
    \end{tabular}
\end{table}

For problem 3NTN, the original PDB structure is a homotrimer. It consists of three helices, which together form a binding site for $\text{Ni}^{2+}$ ion and a binding site for $\text{Cl}^{-}$ ion. When setting up this multi-motif scaffolding problem, we are interested in whether it is possible to combine two binding sites (formed by multiple chains) into a single-chain protein. One possible reason that Genie 2 fails on this task might be that this problem is not solvable given the current specification.

\clearpage

\section{Additional Results on Unconditional Protein Generation}
\label{app:uncond_result}

\subsection{Length-based performance analysis using scTM}

\begin{figure}[h!]
    \centering
    \includegraphics[width=0.96\textwidth]{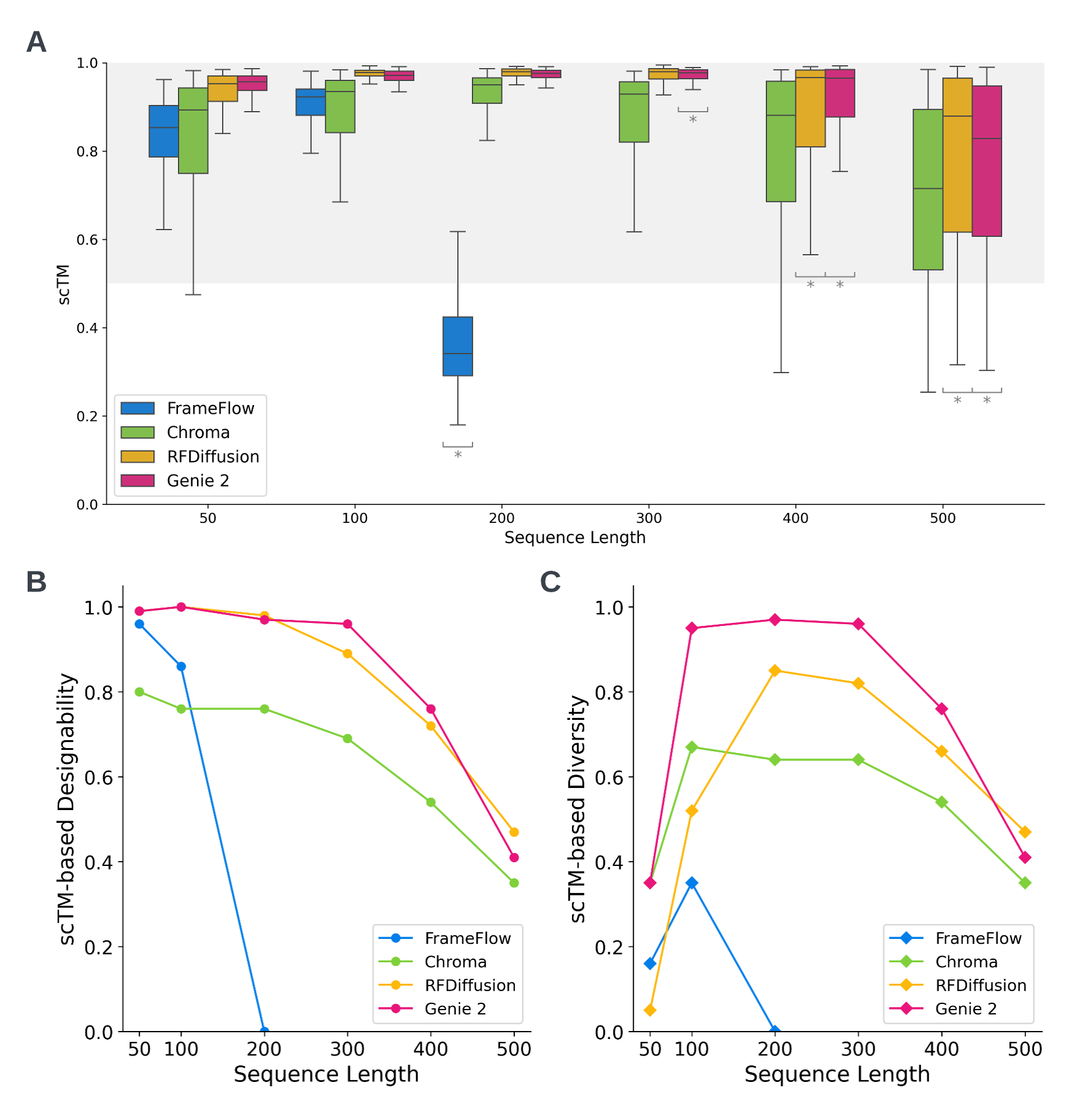}
    \caption{Assessment of Genie 2 and competing methods by sequence length using scTM as the designability metric. 100 structures are generated per sequence length and method. (A) Distribution of self-consistency TM between generated structures and the most similar ESMFold-predicted structures. Asterisk (*) indicates that the sampled sequence length is beyond the maximum sequence length sampled at training time. (B) Plot of scTM-based designability (percentage of scTM-designable structures) as a function of sequence length. (C) Plot of scTM-based diversity (percentage of unique scTM-based designable clusters) as a function of sequence length.}
    \label{appfig:sctm_length}
\end{figure}

We provide additional assessments of Genie 2 and competing methods using a second designability metric, the self-consistency TM score (scTM). scTM is computed using the same pipeline as scRMSD, described in Section \ref{uncond_eval_metric}, except using TM score to measure the structural distance between a generated structure and its most similar ESMFold-predicted structure. scTM is a less stringent metric than scRMSD since TM score is less sensitive to minor structural variations. Figure \ref{appfig:sctm_length}A visualizes the distribution of scTM by sequence length for Genie 2 and competing methods, while Figures \ref{appfig:sctm_length}B and \ref{appfig:sctm_length}C visualize scTM-based designability and diversity as a function of sequence length, respectively. Here, a structure is considered as scTM-based designable if it satisfies both $\text{scTM} > 0.5$ and $\text{pLDDT} > 70$. Diversity is computed using the same clustering procedure described in section \ref{uncond_eval_metric}. Overall trends remain consistent with our main results.

\clearpage

\subsection{Additional examples of designable clusters by Genie 2}

\begin{figure}[h!]
    \centering
    \includegraphics[width=.99\linewidth]{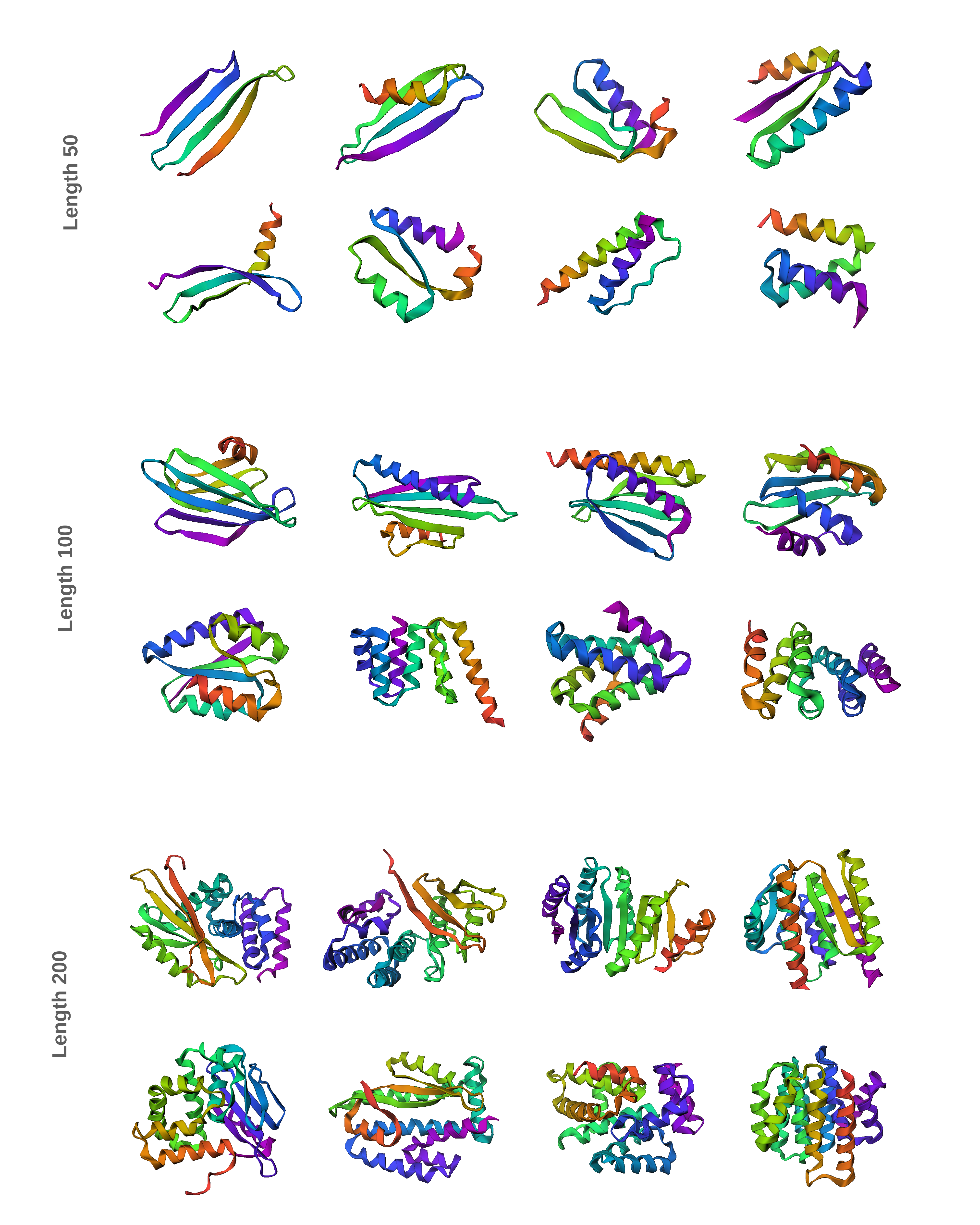}
    \caption{Examples of Genie 2 designed structures with in-distribution sequence lengths (within the maximum sequence length of 256 set at training time).}
    \label{fig:example50_200}
\end{figure}

\begin{figure}[h!]
    \centering
    \includegraphics[width=.99\linewidth]{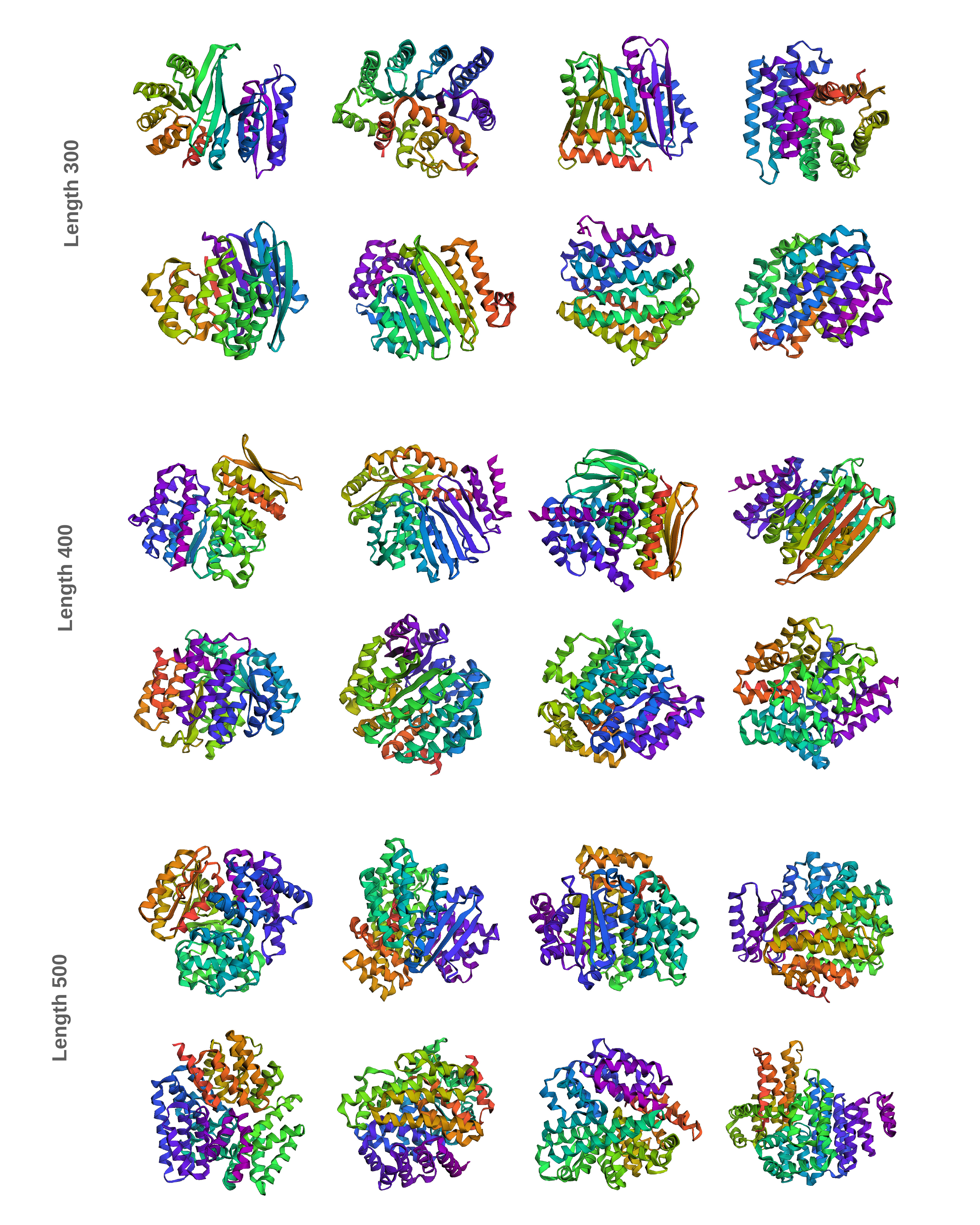}
    \caption{Examples of Genie 2 designed structures with out-of-distribution sequence lengths.}
    \label{fig:example300_500}
\end{figure}

\clearpage

\section{Additional Results on Single-Motif Scaffolding}

\subsection{Evaluation details}

\begin{figure}[h!]
    \centering
    \includegraphics[width=\textwidth]{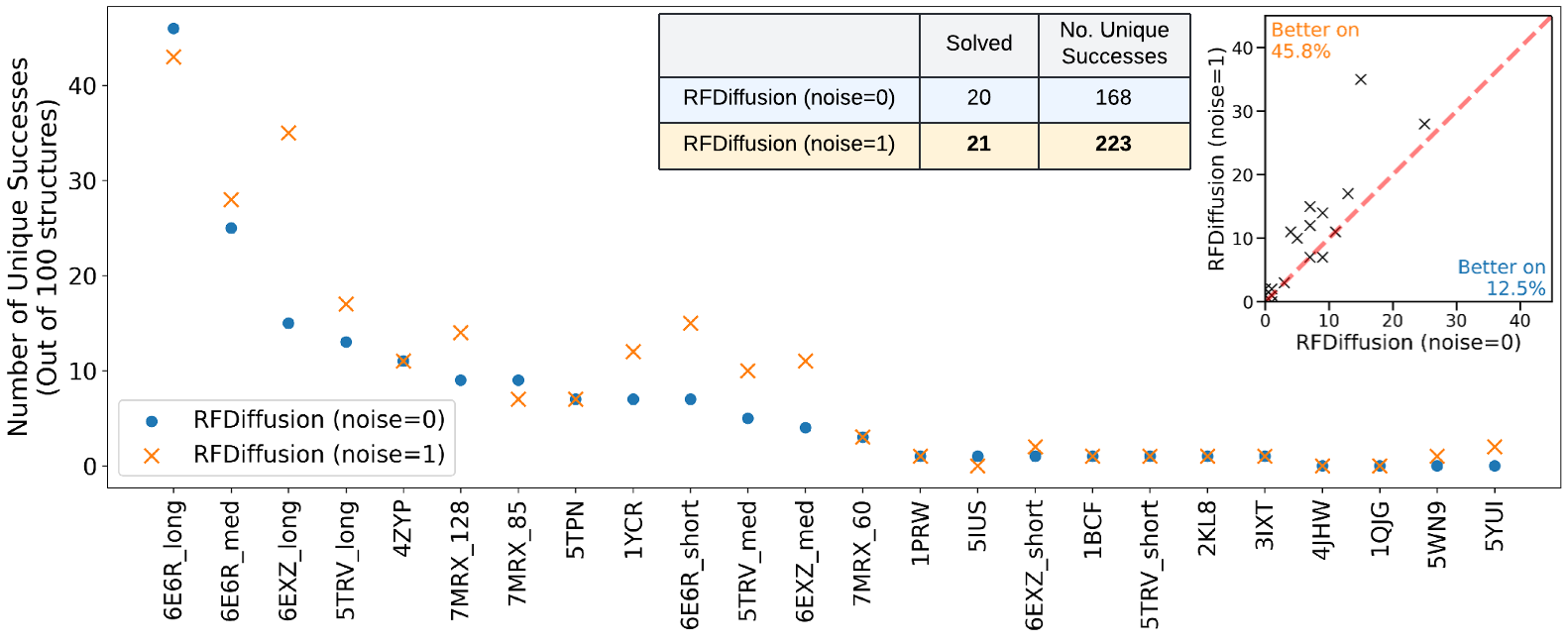}
    \caption{Performance of RFDiffusion with a noise scale of 0 and 1 across 24 single-motif scaffolding tasks. Inset (top right) shows a scatter plot of the (unique) success rate of RFDiffusion with a noise scale of 1 versus RFDiffusion with a noise scale of 0; each point represents a scaffolding task. Summary statistics are shown in table (left).}
    \label{appfig:rfd_noise}
\end{figure}

\citet{watson2023novo} asserts that RFDiffusion achieves a higher success rate when the noise scale is set to 0; however, this success rate does not account for the diversity of designed structures. To ensure a fair comparison, we first assessed the performance of RFDiffusion with noise scale set to 0 and 1. For each motif scaffolding problem, we sampled 100 structures per problem and evaluated them using the same pipeline as in Section \ref{sec:single_motif}. Figure \ref{appfig:rfd_noise} visualizes the number of unique successes by motif scaffolding problems. We observe that RFDiffusion solves more motif scaffolding problems with more diverse designs when the noise scale is set to 1. Thus, to maximize RFDiffusion's performance, we compare Genie 2 with RFDiffusion with a noise scale of 1 throughout this work.

\clearpage

\subsection{Number of unique successes}

\begin{table}[h]
    \caption{Number of unique successes (out of 1,000 structures) generated by Genie 2 and RFDiffusion on each single-motif scaffolding task.}
    \centering
    \setlength{\tabcolsep}{15pt}
    \small
    \begin{tabular}{l|c|c}
        \toprule
         Name & Genie 2 & RFDiffusion \\
         \midrule
        6E6R\_long & \textbf{415} & 381 \\
        6EXZ\_long & \textbf{326} & 167 \\
        6E6R\_med & \textbf{272} & 151 \\
        1YCR & \textbf{134} & 7 \\
        5TRV\_long & \textbf{97} & 23 \\
        6EXZ\_med & \textbf{54} & 25 \\
        7MRX\_128 & 27 & \textbf{66} \\
        6E6R\_short & \textbf{26} & 23 \\
        5TRV\_med & \textbf{23} & 10 \\
        7MRX\_85 & \textbf{23} & 13 \\
        3IXT & \textbf{14} & 3 \\
        5TPN & \textbf{8} & 5 \\
        7MRX\_60 & \textbf{5} & 1 \\
        1QJG & \textbf{5} & 1 \\
        5TRV\_short & \textbf{3} & 1 \\
        5YUI & \textbf{3} & 1 \\
        4ZYP & 3 & \textbf{6} \\
        6EXZ\_short & \textbf{2} & 1 \\
        1PRW & \textbf{1} & \textbf{1} \\
        5IUS & \textbf{1} & \textbf{1} \\
        1BCF & \textbf{1} & \textbf{1} \\
        5WN9 & \textbf{1} & 0 \\
        2KL8 & \textbf{1} & \textbf{1} \\
        4JHW & 0 & 0 \\
        \bottomrule
    \end{tabular}
\end{table}

\clearpage

\subsection{Performance as a function of sample size}
\label{app:single_result_sample_size}

\begin{figure}[h!]
    \centering
    \includegraphics[width=\textwidth]{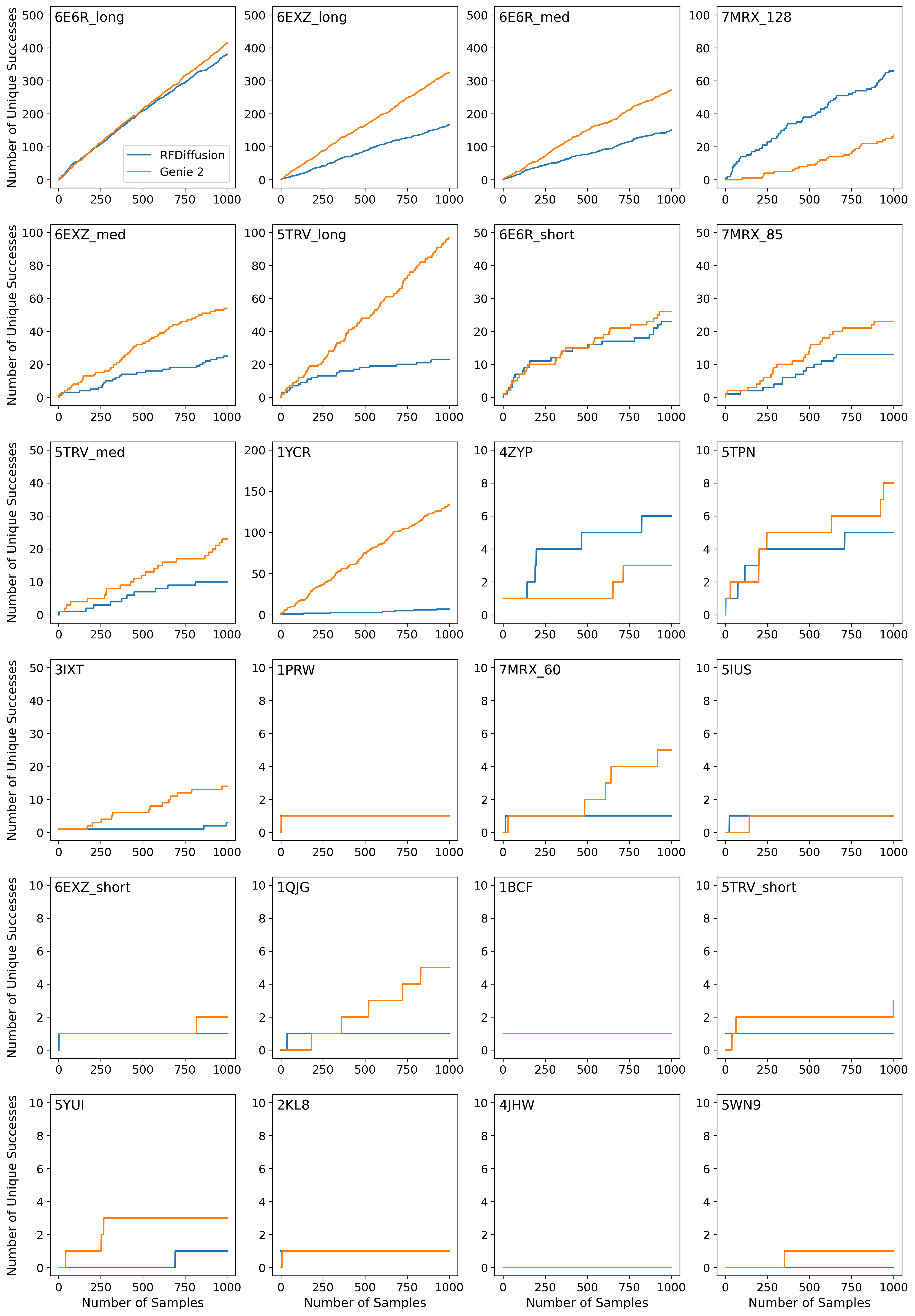}
    \caption{Single-motif scaffolding performance as a function of sample size by problem. The y-axis represents the number of unique successes and is rescaled for each problem.}
    \label{appfig:each_vs_num_samples}
\end{figure}

\clearpage

\subsection{Scatterplot of scRMSD versus motif backbone RMSD}

\begin{figure}[h!]
    \centering
    \includegraphics[width=0.98\textwidth]{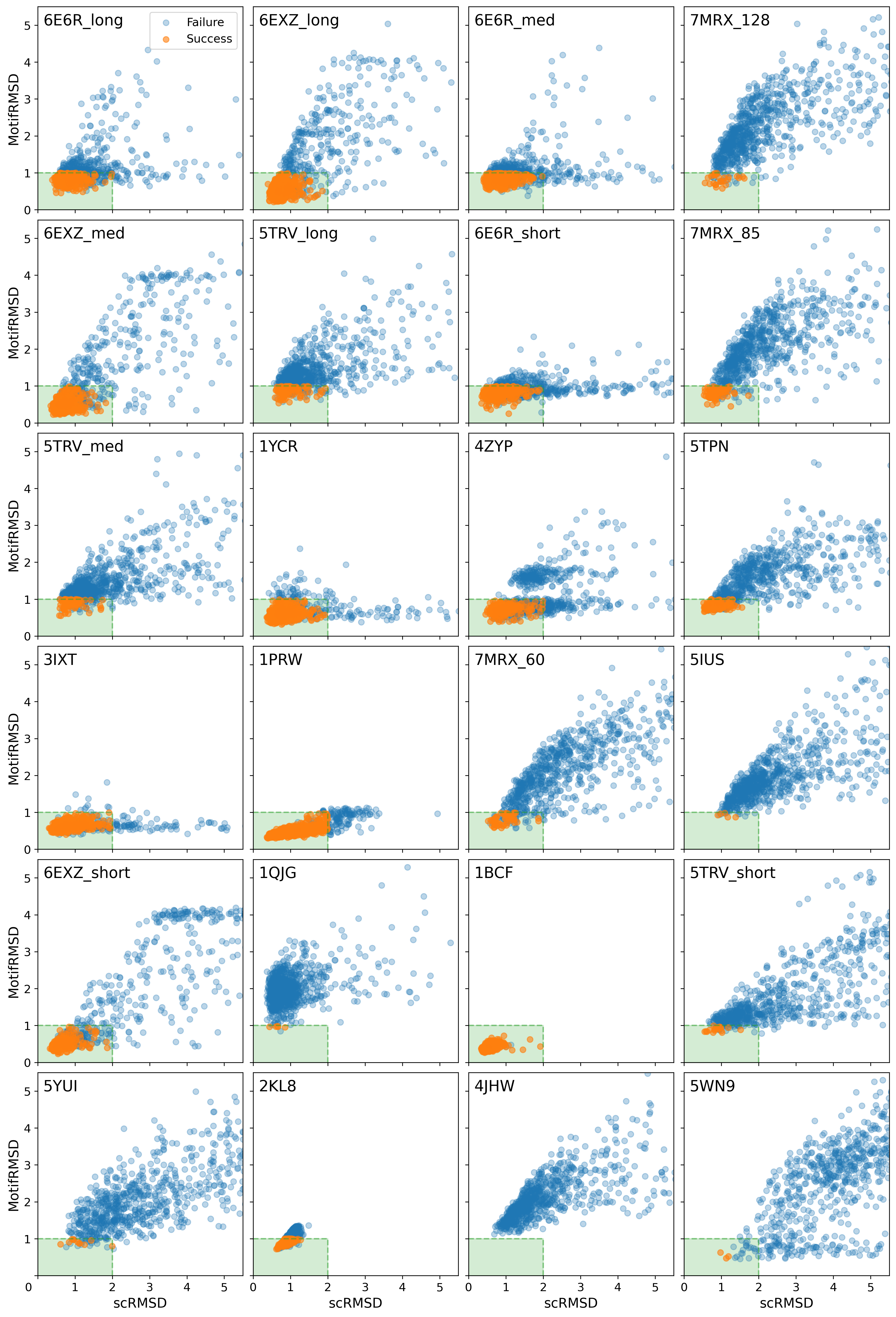}
    \caption{Scatterplot of scRMSD versus motif backbone RMSD by problem, where each point represents one generated structure. The green box denotes the region with $\text{scRMSD} \leq 2$Å and motif backbone $\text{RMSD} \leq 1$Å, which are the two deciding factors of a design's success.}
\end{figure}

\clearpage

\subsection{Additional examples of successful motif scaffolding designs by Genie 2}
\label{app:single_result_examples}

\begin{figure}[h!]
    \centering
    \includegraphics[width=.99\linewidth]{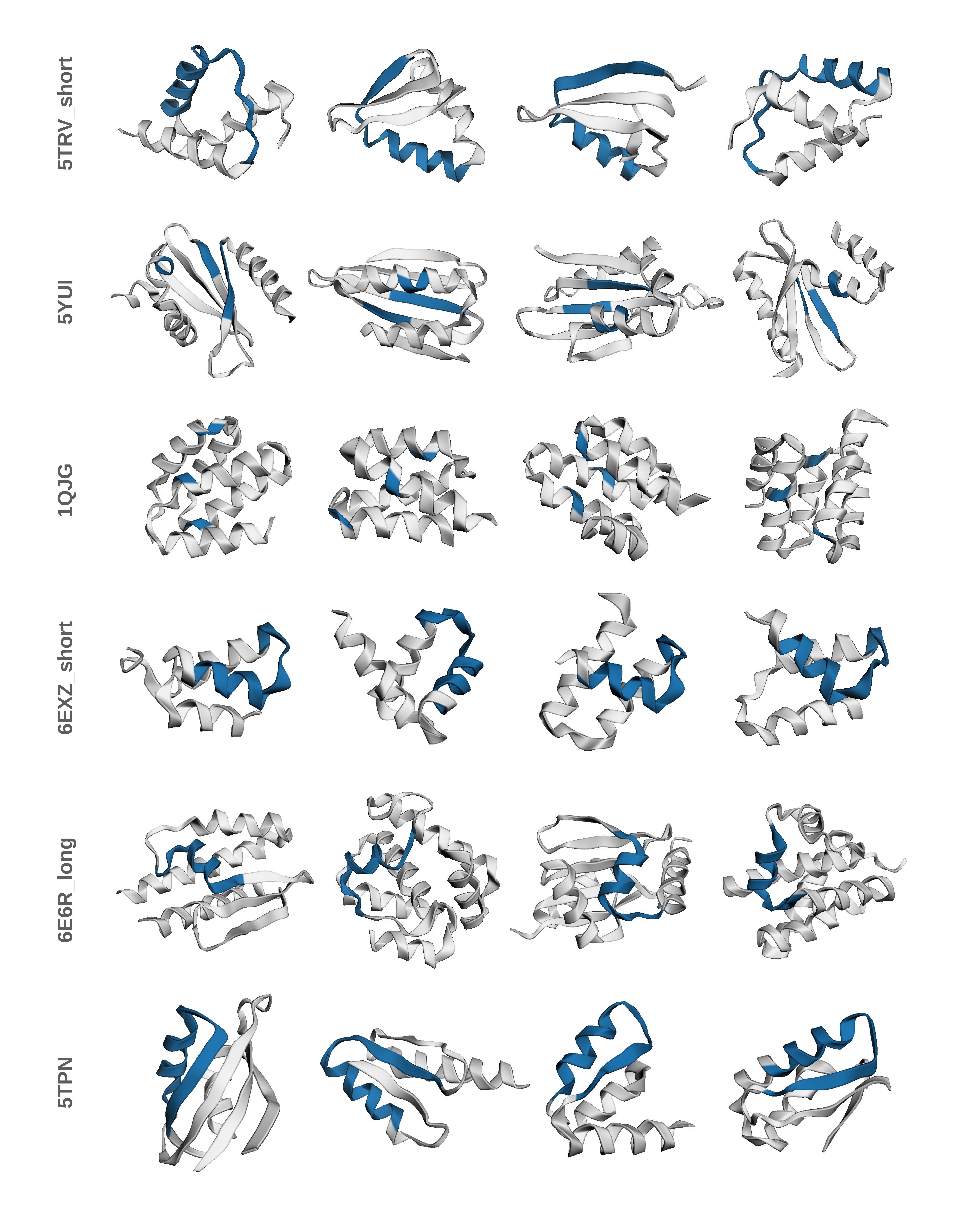}
    \caption{Examples of successfully designed structures by Genie 2 for six single-motif scaffolding tasks. Scaffolds (white) and motifs (blue) are overlaid.}
    \label{fig:example_singlemotif}
\end{figure}

\clearpage

\section{Additional Results on Multi-Motif Scaffolding}
\label{app:multimotif}

\subsection{Number of unique successes}

\begin{table}[h!]
    \caption{Number of unique successes (out of 1,000 structures) generated by Genie 2 on each multi-motif scaffolding task.}
    \centering
    \setlength{\tabcolsep}{15pt}
    \small
    \begin{tabular}{l|c}
        \toprule
         Name & Number of Unique Successes \\
         \midrule
         3BIK+3BP5 & 17 \\
         1PRW\_four & 11 \\
         1PRW\_two & 8 \\
         4JHW+5WN9 & 4 \\
         2B5I & 0 \\
         3NTN & 0 \\
        \bottomrule
    \end{tabular}
    \label{tbl:multimotif_result}
\end{table}

\subsection{Additional examples of successful designs by Genie 2}

\begin{figure}[h]
    \centering
    \includegraphics[width=.99\linewidth]{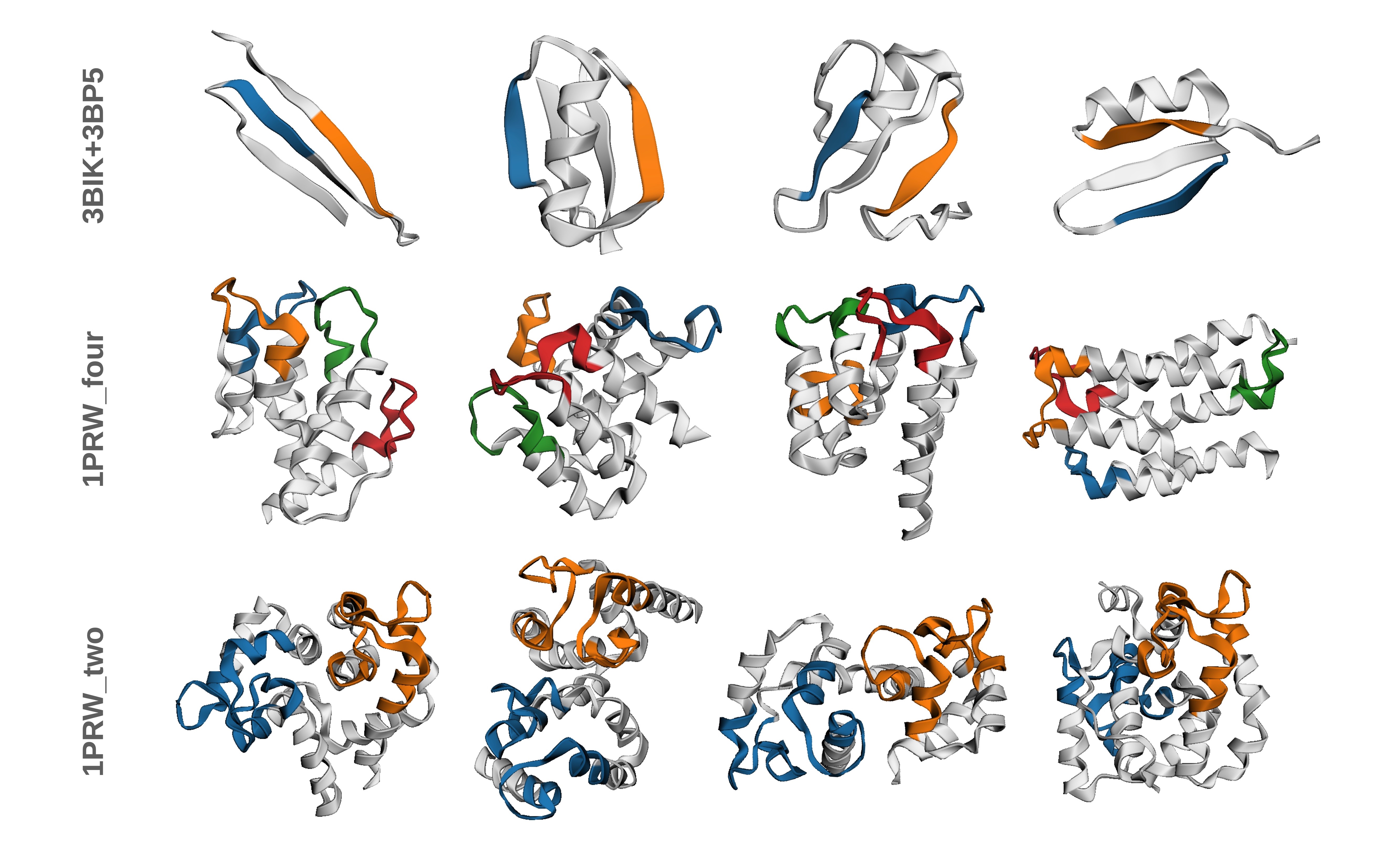}
    \caption{Examples of successfully designed structures by Genie 2 for three multi-motif scaffolding tasks. Scaffolds are in grey and different motifs are colored differently. For 4JHW+4WN9, all four unique successes are shown in Figure~\ref{fig:multiscaffoldresults}.}
    \label{fig:example_multi}
\end{figure}

Figure~\ref{fig:example_multi} shows the successful designs of three multi-motif scaffolding problems. The designs for 3BIK+3BP5 exhibit diverse secondary structures, including structures containing strands (first), helices (second and fourth), and loops (third). For tasks 1PRW\_two and 1PRW\_four, the loops of EF-hand motifs that interact with substrates are well exposed to the surface in all designs. The structures are diverse and clearly different from the original 1PRW, with the EF-hand motifs distributed asymmetrically throughout the structure. In the designs of 1PRW\_two, the 4-helix bundles are also in different relative orientations compared to the original 1PRW. For example, in the second design, the loops of two bundles face the same side. These diverse and novel designs open the possibility of creating more stable or functional proteins with the desired motifs.

\clearpage

\section{Sampling Time}
\label{app:sample_time}

In this section, we compare the generation times of Genie 2, RFDiffusion, FrameFlow, and Chroma at different lengths. We use a single A6000 GPU (48GB memory) and average the inference time of a single sample over 10 runs. For RFDiffusion, we use the self-conditioning sampler and exclude pLDDT and amino acid prediction for fair comparison. For Chroma, we use unconditional monomer sampling and exclude amino acid prediction. We use the simple profiler from PyTorch Lightning \cite{Falcon_PyTorch_Lightning_2019} to profile inference function calls of FrameFlow and Genie.
 
\begin{table}[h]
    \centering
    \caption{Sampling time of different methods for proteins of different lengths.}
    \begin{tabular}{lcc|cccccc}
    \toprule
    \multirow{2}{*}{Method} & \multirow{2}{*}{Parameters} & \multirow{2}{*}{Timesteps} & \multicolumn{6}{c}{Length} \\
    & & & 50 & 100 & 200 & 300 & 400 & 500 \\
    \midrule 
    Genie 2 & 15.7M & 1000 & 42.2 & 49.4 & 71.4 & 134 & 236 & 353 \\
    RFDiffusion & 59.8M & 50 & 18.7 & 21.4 & 41.2 & 80.1 & 137 & 214\\
    FrameFlow & 17.4M & 100 & 3.45 & 4.33 & 6.50 & 9.84 & 13.8 & 18.5 \\
    Chroma & 18.5M & 500 & 22.0 & 22.6 & 29.0 & 35.4 & 41.8 & 48.5 \\
    \bottomrule
    \end{tabular}
    \label{tbl:appSamplingTime}
\end{table}

\clearpage  

\end{document}